\documentclass[a4paper,11pt]{article}
\usepackage[utf8]{inputenc}
\usepackage[margin=2.5cm]{geometry}
\usepackage{graphicx}
\usepackage{epsfig}
\usepackage{amsmath}
\usepackage{amssymb}
\usepackage{cite}
\usepackage{float}
\usepackage[font={footnotesize,it}]{caption}

\begin{document}

\begin{titlepage}
\title{Rotating BTZ Black Holes and One Dimensional Holographic Superconductors}
\author{}
\date{
Pankaj Chaturvedi and Gautam Sengupta
\thanks{\noindent E-mail:~ cpankaj@iitk.ac.in , sengupta @iitk.ac.in}
\vskip0.4cm
{\sl Department of Physics, \\
Indian Institute of Technology,\\
Kanpur 208016, \\
India}}
\maketitle
\abstract{
\noindent
We consider charged rotating BTZ black holes in 2+1 dimensions and obtain 1+1 dimensional holographic superconductors on a spatial circle in the context of the $AdS_3/CFT_2$ correspondence. The charged condensate for the boundary superconductor is computed both in the analytic and the numerical framework in a probe limit and a low angular momentum approximation.  A critical value of the angular momentum for the onset of superconductivity is established. We also numerically compute the electrical conductivity of the 1+1 dimensional boundary theory on a circle. The conductivity exhibits a dependence on angular momentum of the rotating black hole both for the normal and the superconducting phase of the boundary field theory.  The significance of the boundary field theory in the context of a Fermi-Luttinger liquid on a circle is discussed.}

\end{titlepage}

\section{Introduction}

One of the most significant insights in fundamental physics within the last decade is undoubtedly the gauge gravity correspondence. This relates a weakly coupled theory of gravity in a bulk AdS space time to a strongly coupled conformal field theory on the asymptotic boundary of the AdS space-time, in the large N limit and vice versa and has been hence described as  a {\it holographic duality }. For a black hole in the bulk AdS space time it could be shown that the corresponding boundary theory was at a finite temperature equal to the Hawking temperature of the black hole. In recent years there has been an intense focus in the investigation of such boundary theories at finite temperature and finite chemical potential  that involves the exciting possibility of describing the properties of strongly coupled condensed matter systems. One of the main developments in this context has been the construction of holographic superconductors as boundary theories through the gauge gravity correspondence. It was shown by 
Gubser \cite {Gubser} that charged Reissner Nordstrom-AdS ( RN-AdS) black holes in the presence of a charged scalar field were unstable to the formation of ``scalar hair" below a certain critical temperature. Following the AdS-CFT dictionary Hartnoll et al \cite {Hartnoll1,Hartnoll2} demonstrated that this instability in the bulk is translated to the boundary theory as a superconducting instability which leads to the formation of a condensate corresponding to some charged operator $\cal {O}$.  The corresponding superconducting phase of the boundary theory characterized by a charged condensate and zero dc resistivity could be explicitly realized numerically. The essential physics of this phase transition could be understood in the probe limit of a large scalar field charge, in which case the back reaction of the condensate on the bulk geometry may be neglected as a first approximation. The effect of the back reaction could be accounted for in a systematic perturbative computation. The local abelian gauge 
symmetry is broken in the bulk by the scalar hair and this translated to a broken global abelian gauge symmetry in the boundary theory. Thus strictly speaking the boundary theory exhibits superfluidity but the distinction is not significant in the context of conductivity or other transport properties and it may be assumed that the boundary theory is {\it weakly gauged}.  

The explicit realization of  {\it holographic superconductors } inspired an extensive and systematic study of their condensate formation, transport and spectral properties in diverse dimensions \cite {Hartnoll1,Hartnoll2,Horowitz} both in the probe limit and including the back reaction \cite{GregorySoda,Davood,LiuSun,GubserNellore,HorowitzWay,Ammon-et,Brihaye1,BarclayGregory,Siani,GregoryRev,CaiBackreaction}. Further it was shown in \cite{Salvio} that it was possible to include dynamical gauge fields through Neumann type condition at the AdS boundary which influences the superconducting phase transition through vortex formation. The translationally invariant bulk theory leads to a divergence of the Drude peak at zero frequency and a gap formation in the real part of the electrical conductivity which is characteristic of superconducting phase transitions \cite {HartnollRoberts1,HartnollRoberts2,Amado,Nishioka,Herzog}. The analysis was later extended to non abelian gauge fields and tensor fields in the bulk 
leading to p-wave and d-wave superconductors \cite{GubserPufu,Basu,Yanyan1,Yanyan2,BasuMukherjee,HerzogPufu,AmmonKerner1,AmmonKerner2,PeetersZamaklar,ChenYeh,BeniniYarom,BeniniHerzog,ZengZong} . Several other systems in higher dimensions involving the addition of higher curvature terms like the Gauss Bonnet term \cite {Barclay,Sugumi,Setare,Jian} and also superconductors arising from nonlinear Born-Infeld electrodynamics \cite {JingChen,BaiXu} have been studied.  All of these constructions  involved the charged spherically symmetric RN-AdS black holes in the bulk space time. In \cite {Sonner} a four dimensional charged rotating Kerr-Newman AdS black hole was considered as the gravity dual to a 2+1 dimensional rotating holographic superconductor on a two sphere at the boundary in the framework of $AdS_4/CFT_3$. In this case the Lense-Thirring effect \cite {Lense} induced a boundary rotation which was equivalent to an effective magnetic field in the boundary theory. Subsequently in \cite {Brihaye2} a four 
dimensional rotating black string solution in the bulk was demonstrated to be dual to a holographic superconductor on the $S^1 \times R$ boundary. 

Its well known however that  condensed matter physics in 1+1 dimensions involve interesting phenomena such as spin chains, quantum wires and Luttinger liquids which provides a natural motivation to investigate lower dimensional boundary field theories in the context of holographic superconductors. The consequent gravity duals for such boundary theories in lower dimensions exhibit a rich and interesting variety. One of the most exciting avenues in this context is the study of  1+1 dimensional boundary theories in the context of  the $AdS_3/CFT_2$ correspondence. The dual theory of gravity in the bulk in this case is often considered to be a BTZ black hole in 2+1 dimensions but other gravity duals have also been considered. In \cite{Nilanjan} a charged BTZ black hole in the bulk was considered to study 1+1 dimensional boundary theories with a background electric charge. Bulk fermions in such a gravity background were shown to lead to a boundary field theory, certain phases of  which resembled Fermi-Luttinger 
liquids \cite{Bala,Aninda}. It was shown in \cite {Ren} that a charged BTZ black hole in the presence of  a charged scalar field in the bulk leads to a 1+1 dimensional holographic superconductor at the boundary  \footnote {Note that the bulk theory considered was a pure Einstein-Maxwell theory in 2+1 dimensions without Chern Simons gauge fields the inclusion of which does not lead to any bulk instability. } \cite {KJensen, Rleigh}. In \cite {Iqbal} the effect of bulk magnetic monopole tunneling events on the density density correlations is studied for a $(2+1)$ dimensional Maxwell-Einstein bulk which leads to Friedel oscillations in the (1+1) dimensional boundary theory.

In the context of  $AdS_3/CFT_2$ it was shown in \cite{Bala} that fermions in a 2+1 dimensional rotating BTZ black hole with Chern Simons gauge fields and Wilson lines in the bulk was dual to a helical Luttinger liquid on the  1+1 dimensional boundary.
Later following \cite {Ren} it was shown in \cite {Nurma} that a charged rotating BTZ black hole in the presence of  a charged scalar field leads to a 1+1 dimensional holographic superconductor at the boundary, in a low angular momentum approximation and the probe limit. However the analytic treatment of the rotating case in \cite {Nurma} appears to have several lacunae and lacks a clear perspective on the interesting 1+1 dimensional boundary theory. There also seems to be several incorrect analytic expressions in the treatment and fails to indicate that the boundary theory is actually on a spatial circle. In fact a computation of the conductivity of this interesting 1+1 dimensional boundary theory is not even attempted in \cite {Nurma}. Our motivation is to comprehensively investigate the instability of a charged rotating BTZ black hole in the presence of a charged scalar field in the bulk and study the conductivity of the 1+1 dimensional boundary field theory on a circle. To this end we analytically 
establish the condensate formation in the boundary theory addressing the lacunae in \cite {Nurma} and arrive at the correct expression for the condensate in the probe limit and the low angular momentum approximation. We further compute the condensate formation numerically and obtain graphical plots which are compared with the plots obtained from the analytic treatment. We then obtain  the ac conductivity for the strongly coupled  1+1 dimensional boundary theory in the framework of  the linear response theory, both for the normal and the superconducting phase. It is observed that the ac conductivity in the strongly coupled boundary theory exhibits a dependence on the angular momentum of the rotating black hole.

The article is organized as follows, in Section 2 we briefly outline and collect the main results for the setup of  the AdS$_3$-CFT$_2$ correspondence and follow that up with a discussion of the non rotating charged BTZ black hole in the probe limit and the associated holographic superconductor in Section 3. In Section 4 we present both the analytic and numerical computations for the condensate formation and ac conductivity in the 1 +1 dimensional boundary theory corresponding to a bulk charged rotating BTZ black hole in the probe limit and low angular momentum approximation. In Section 5 we present a summary of our results and discussions. 

\section{ Holographic setup for $AdS_{3}/CFT_{2}$ }

The AdS/CFT duality relates a bulk field theory in an AdS space time with a field theory residing at the asymptotic boundary of AdS space. If the boundary theory is at its strong coupling limit then it is dual to a bulk theory which is purely gravitational. The recipe for obtaining the boundary field theory correlators from gravity computations is referred to as GKPW \cite {GKPW,Witten} prescription. According to this prescription one equates the partition function of the bulk theory taken to be a functional of the boundary values of the bulk fields, to the generating functional for the correlators of the boundary field theory. The prescription maybe stated as follows,
\begin{equation}
{\cal Z}_{grav}[\phi_{0}]= <e^{-\int \phi_{0}{\cal O}} >_{CFT}
\end{equation}
Here $\phi_0=\phi_0(x)$ is the boundary value of the bulk field $\phi(x)$ and the right hand side of the above equation is the generating functional of the boundary CFT for the boundary operator $\cal O$ dual to the bulk field $\phi(x)$. The correlators for the boundary CFT may be expressed as
\begin{equation}
<{\cal O}(x_1){\cal O}(x_2).....{\cal O}(x_n)> = \frac{\delta}{\delta \phi_0(x_1)}\frac{\delta }{\delta \phi_0(x_2)}....\frac{\delta}{\delta \phi_0(x_n)}S_{grav}^{(onshell)}\vert_{\phi_0 =0} ,\label{eq:corrGKPW}
\end{equation} 
where $S^{(onshell)}_{grav}$ is the extremum of  the gravitational action evaluated at the boundary with appropriate boundary conditions. It is observed from the solutions to the bulk equations of motion that the asymptotic behaviour near the boundary for any field $\phi$ propagating in the bulk AdS space time is given as
\begin{equation}
 \phi(z)= {\cal A} z^{\Delta_{-}} (1+\cdots)+{\cal B} z^{\Delta_{+}}(1+\cdots), \label{eq:adsfield}
\end{equation}
where the dots represent the regular terms which vanish in the limit $z \rightarrow 0$. The characteristic exponents $\Delta_{\pm}(\Delta_{-} < \Delta_{+})$ may be evaluated from the perturbation equations for the field. For example a scalar field is described by the exponents $\Delta(\Delta-d)=m^{2}L^{2}$ whereas for vector fields they are given as $\Delta(\Delta-d+2)=m^{2}L^{2}$.  Near the boundary the first term given in \cite {Ren} is dominant so the quantity $\cal A$ is taken to be the source for an operator $\cal O$ dual to the field $\phi$ while the quantity $\cal B$ is treated as the expectation value  $<\cal O>$ of the operator. The retarded Green function at the boundary with incoming wave boundary conditions at the horizon may be obtained from the  GKPW prescription and is expressed as $<\cal O O>_{R}\sim {\cal B}/{\cal A} $. The nature of the solution at the boundary also depends on the quantity  $\nu=\frac{\Delta_{+}-\Delta_{-}}{2}=\sqrt{(d/2)^{2}+m^{2}L^{2}}$. For integral values 
of $\nu$ the solution contains a logarithmic term which is absent for non-integral values. 

The transport properties of the boundary field theory like the ac conductivity may be extracted from the Green function (obtained by the GKPW prescription) through the usual Kubo formula of  the linear response theory in the long wavelength and low frequency limit. In this framework a conserved current density $J_{i}$ is proportional to the external vector potential $A_{j}$ such that, $J_{i}=G^{ret}_{ij}A^{j}$, where $G^{ret}_{ij}$ is the retarded Green function and is given as
\begin{equation}
G^{ret}_{ik}(x-x',t-t')=-i\theta(t-t')<[J_{i}(x,t),J_{k}(x',t')]>
\end{equation}
The current-current correlators may be obtained from (\ref{eq:corrGKPW}) with the bulk field $\phi(x)$ as the gauge field $A_{j}(x)$ while the corresponding operator $\cal O$ in the boundary theory as the current $J_{i}$.

Having described the general setting for the  AdS-CFT correspondence we now proceed to outline the essential holographic dictionary for the AdS$_3$/CFT$_2$ correspondence in the context of a charged BTZ black hole in the bulk and the corresponding 1+1 dimensional boundary field theory. The bulk theory is described by a Einstein-Maxwell action coupled to a charged scalar field in 2+1 dimensions and is given as
\begin{equation}
S=\int d^{3}x\sqrt{-g}\left(R+\frac{2}{L^2}-\frac{1}{4}F_{\mu\nu}F^{\mu\nu}-|\nabla\Psi-iqA\Psi|^2-V(\Psi)\right),\label{eq:action}
\end{equation}
where, $V(\Psi)=m^2|\Psi|^2$ or it can be a derivative of other exotic terms made up of $\Psi$ \cite{FrancoGarcia}. One starts with the following metric ansatz for AdS$_3$ in Poincare coordinates
\begin{equation}
ds^{2}=\frac{L^{2}}{z^{2}}(-f(z)dt^{2}+dx^{2}+\frac{dz^{2}}{f(z)}),\label{eq:ansatz}
\end{equation}
here $z$ describes the dimension into the bulk and the other coordinates parametrize the 1+1 dimensional boundary with $L$ descrbing the AdS length scale. It is evident that the conformal boundary is at $z=0$ and the horizon at $z=z_{h}$ such that $f(z_{h})=0$. The metric has two scaling symmetries \cite {Hartnoll1,Hartnoll2} which may be used to set $L=1$ and $z_{h}=1$.

The conductivity for the 1+1 dimensional boundary field theory in our case is obtained by adding a perturbation $e^{-i\omega t}A_x(z)$ to the system and then solving the linearized equations in the bulk for $A_x$. The solution to the linearized equation for $A_x$ with the incoming boundary wave condition at the horizon yields the retarded Green function. The Green function  is given as $G^{xx}=J^x/A_x$, where $x$ denotes the spatial dimension and $J_x$ is the conserved current that measures the linear response with respect to perturbations of the vector potential $A_x$. Since the current density $J^x=\sigma^{xx}E_x=i\omega\sigma^{xx}A_x$, this leads to the expression for the ac conductivity is $\sigma^{xx}=G^{xx}/i\omega$.

 
\section{Boundary Theory for the Charged BTZ Black Hole}

In this section we will briefly outline the properties of  the 1+1 dimensional boundary field theory dual to the bulk charged BTZ black hole in the presence of  a charged scalar field in the probe limit \cite {Ren}. In the context of the $AdS_{3}/ CFT_{2}$  duality a non zero profile for the charged scalar field in the bulk corresponds to the condensate of an operator ${\cal O}$ in the boundary theory which leads to the superconducting phase transition. A solution to the equations of motion following from the 2+1 dimensional bulk action as given in (\ref {eq:action}) is described by the following expressions for the metric and the gauge field  

\begin{equation}
 ds^{2} = \frac{1}{z^{2}}(-f(z)dt^{2}+dx^{2}+\frac{dz^{2}}{f(z)}),\label{eq:red}
\end{equation}
\begin{equation}
f(z)=1-\frac{z^{2}}{z_{h}^{2}}+\frac{\mu^{2}z^{2}}{2}\ln{\frac{z}{z_{h}}} 
\end{equation}
 \begin{equation}
 A(z)=\mu\ln{\frac{z}{z_{h}}}dt
\end{equation}

This solution corresponds to a charged BTZ black hole and in the limit of $z \rightarrow 0$ at the boundary the metric is identical to pure $ AdS_{3}$ with $z=z_{h}=1$ is the horizon. The Hawking temperature $T_{h}$ of the black hole is given as $T_{h}=\frac{|f'(1)|}{4\pi}=\frac{4-\mu^{2}}{8\pi}$. Notice that the temperature depends on the chemical potential $\mu$ that provides a scale in the theory which drives the phase transition. The normal phase of the boundary theory is described by the charged BTZ black hole at a temperature $ T \ >\ T_c$ with a vanishing profile for the scalar field while the superconducting phase is given by the charged BTZ black hole with scalar hair at a temperature $ T \ <\ T_c$.

\subsection{Normal Phase}

As described in the above section the normal phase of  the 1 + 1 dimensional holographic superconductor is dual to a charged BTZ black hole without scalar hair. This corresponds to the solution $\Psi=0$ and $A_t(z)=\phi(z)= \mu\ln(z/z_h)$ for the equations of motion following from the action (\ref{eq:action}). To compute the ac conductivity for the normal phase of the boundary theory including the backreaction of the gauge field, one perturbs the bulk configuration through the perturbations $e^{-i\omega t}A_x(z)$ for the vector potential and $e^{-i\omega t}g_{tx}(z)$  for the metric that leads to two linearized equations for $A_x$ and $g_{tx}$ in the background given by (\ref{eq:ansatz}). Eliminating the metric perturbation $g_{tx}$, a single equation for $A_x$ may be obtained as
\begin{equation}
A_x''+\left(\frac{f'}{f}+\frac{1}{z}\right)A_x'+\left(\frac{\omega^2}{f^2}-\frac{A_t'^2z^2}{f}\right)A_x=0,\label{eq:pertAx}
\end{equation}
here the prime denote derivative with respect to $z$ and from the above equation we can see that the near boundary behavior of $A_x$ is given as 
\begin{equation}
A_x ={\cal A}\ln(z)+{\cal B}+\cdots,\label{eq:bdryAx}
\end{equation}
where the dots are the regular terms that vanish in limit $z\to 0$. We see that near the boundary the leading term is ${\cal A}\ln z$, and this identifies  ${\cal A}$ to be the source \footnote { There is some controversy regarding this choice, we follow the choice described in \cite {Nilanjan}}. So the Green function is given as \cite {Ren, Son}

\begin{equation}
G=-\frac{\cal B}{\cal A}.\label{eq:green}
\end{equation}

From (\ref {eq:pertAx}) we observe that the solution for  $A_{x}$ near the boundary is given by (\ref{eq:bdryAx}), while the near horizon form of  $A_{x}$ with the incoming wave boundary condition is given as
\begin{equation}
A_{x}(z)|_{z \rightarrow 1} = (1- z^{2})^{2i\omega/(\mu^{2} - 4)}(1+\cdots) .
\end{equation}
Using the equations (\ref{eq:green}) and (\ref{eq:bdryAx}) the expression for the ac conductivity 
in terms of  the boundary value of the field $A_{x}$ for a small cutoff  $z= \epsilon$ near the horizon may be expressed as
\begin{equation}
\sigma(\omega) = \frac{i(A_{x}- z A'_{x} z \ln{z})}{\omega (z A'_{x})}|_{z \rightarrow \epsilon}.
\end{equation}
In \cite {Ren,Nilanjan} the behavior of the real and imaginary parts of the ac conductivity were studied numerically. Their results show that the real part of conductivity decays exponentially with a delta function near $\omega=0$, which corresponds to a pole in the imaginary part of the conductivity. The dc limit of the real part of the conductivity ie. $Re(\sigma_{dc})$ decreases with temperature. At zero temperature when the black hole is extremal the conductivity vanishes.

\subsection{Superconducting Phase} 

The superconducting phase of  the boundary theory is described by a charged BTZ black hole with scalar hair which occurs below a certain critical temperature $T_c$. In the probe limit where the back reaction of the gauge field and the scalar field on the bulk metric is neglected  the lapse function is given as $f(z)=1-z^2$ \cite {Ren}. The equations of motion for the bulk fields in the probe limit  are hence given as

\begin{eqnarray}
R_{mn}-\frac{1}{2} g_{mn}(R+\frac{2}{L^2})=0,\label{EinsAdS}\nonumber\\
\frac{1}{\sqrt{-g}}D_{m}(\sqrt{-g}D^{m}\Psi)-m^{2}\Psi=0 \nonumber\\
\partial_{n}(\sqrt{-g}F^{nm})+ i \sqrt{-g}(\Psi\overline{D}^{m}\Psi^{*}-\Psi^{*}D^{m} \Psi)=0,\label{eq:bulkeom}
\end{eqnarray}
The equations of motion in (\ref {eq:bulkeom}) with the ansatz $\Psi=\psi(z)$, $A_{t}=\phi(z)$ maybe expressed as follows
\begin{eqnarray}
\psi''+\left(\frac{f'}{f}-\frac{1}{z}\right)\psi'+\frac{\phi^2}{f^2}\psi-\frac{m^2}{z^2f}\psi &=& 0,\\
\phi''+\frac{1}{z}\phi'-\frac{2\psi^2}{z^2f}\phi &=& 0,\\
A_{x}''+\left(\frac{f'}{f}+\frac{1}{z}\right)A_{x}'+\left(\frac{\omega^2}{f^2}-\frac{z^2\phi'^2}{f}-\frac{2\psi^2}{z^2f}\right)A_{x} &=&0,\label{eq:EOM}
\end{eqnarray}
where the last equation is linearized in $A_x$. From the above equations the form of the solutions near the boundary may be obtained as

\begin{eqnarray}
\psi &=& \psi_{1}z\ln z+\psi_{2}z+\cdots,\label{eq:psib}\\
\phi &=& \mu\ln z+\rho+\cdots,\label{eq:phib}\\
A_{x} &=& {\cal A}\ln z+{\cal B}+\cdots,\label{eq:bsol}
\end{eqnarray}

The boundary conditions for the bulk fields at the horizon $z=1$ are 

\begin{equation}
\phi(z)\vert_{z=1}=0,~~~ \psi(z)= 2 \psi'(z)\vert_{z=1} ,~~~~ A_{x}\vert_{z=1}\sim (1-z^2)^{-i\omega/2}+\cdots . \label{eq:bdrycond}
\end{equation}

The retarded Green function as described in \cite{Ren} may be read off from the form of $A_{x}$ near the boundary as 

\begin{equation}
G = - \frac{\cal B}{\cal A}= -\frac{A_{x}-z A'_{x}\ln{z}}{z A'_{x}}\vert_{z \rightarrow \epsilon} \label {eq:retdgf}
\end{equation}

The mass of scalar field near the BF bound  \cite {Breitenlohner} is taken as $m^{2} = m^{2}_{BF}=-1$ in order to examine the condensation of the dual operators ${\cal O}_{1}$ and ${\cal O}_{2}$ corresponding to $\psi_{1}$ and $\psi_{2}$ in the expansion of $\psi$ given by (\ref{eq:psib}). The spontaneous symmetry breaking for the superconducting phase transition, requires that the condensation of the operators should occur without being sourced. Thus for obtaining the superconducting phase we have two different sets of boundary conditions for the two operators. This in fact corresponds to the choice of two distinct statistical ensembles defining the boundary field theory at a finite temperature and a finite chemical potential. The expectation value of the operators for the two different boundary conditions are given as 
\begin{eqnarray}
<{\cal O}_{1}> = \psi_{1},~\psi_{2}=0\\
<{\cal O}_{2}> = \psi_{2},~ \psi_{1}=0
\end{eqnarray}

The detailed study of the formation of the condensates ${\cal O}_{1}$ and ${\cal O}_{2}$ along with their properties are described in \cite {Ren}. Here we simply state their results for the dependence of  both the condensates on the temperature. This is given as follows,
\begin{equation}
<{\cal O}_1>\approx 5.3 \ T_c \ (1-T/T_c)^{1/2},
\end{equation}
where $T_c\approx 0.050 \ \mu$, and
\begin{equation}
<{\cal O}_2>\approx 12.2 \ T_c \ (1-T/T_c)^{1/2},
\end{equation}
where $T_c\approx 0.136 \ \mu$. From \cite{Ren} it is seen that there is a second order phase transition and the real part of the conductivity falls of exponentially with the formation of a gap near $\omega_{g}$. The imaginary part of the conductivity on the other hand has a pole corresponding to a delta function at $\omega=0$ in the real part of the conductivity. Both the real and imaginary parts of the conductivity follow the standard Kramers-Kronig relation and the FGT sum rules. 


\section{Rotating Charged BTZ Black Hole and the Boundary Theory}

In this section we will begin the investigation of the 1+1 dimensional strongly coupled boundary field theory which is dual to a charged rotating BTZ black hole in the bulk AdS$_3$ space time. An attempt was made in \cite {Nurma} to compute the superconducting phase transition in the boundary theory dual to such a bulk background, however as mentioned earlier there were
several lacunae in their computation and a clear perspective on the interesting boundary field theory was missing. To this end we recompute the superconducting phase transition in this theory with the insertion of the correct terms in the expression for the corresponding equations of motion for the bulk fields and obtain a correct expression for the charged condensates. We also further investigate the transport properties like the ac conductivity of the 1+1 dimensional boundary field theory on a circle $S^1$.

In order to study the 1+1 dimensional boundary theory dual to a rotating BTZ black hole in the bulk 
in the context of $AdS_{3}/CFT_{2}$ we begin with the action (\ref{eq:action}). We then consider a solution
to the equations of motion which corresponds to a charged rotating BTZ black hole in the bulk.
In the BTZ coordinates $(r, t, \varphi)$ this may be written down as 
\begin{equation}
ds^{2}=-\frac{r^2}{L^2}f(r)dt^{2}+\frac{L^2}{r^2}\frac{dr^{2}}{f(r)}+r^{2}\left(d\varphi-\frac{J}{2r^2}dt\right)^{2},\label{eq:BTZCR}
\end{equation}
where,
\[f(r)=1-\frac{M L^{2}}{r^{2}}+\frac{J^{2} L^{2}}{4r^{4}}-\frac{\mu^{2}L^{2}}{2r^{2}}\ln{r}.\]
Note that the coordinate $r$ defines the direction into the bulk and the boundary is parametrized by the coordinates $(t,\varphi )$ where $0 \leq \varphi \leq 2\pi$ is an angular coordinate.  Hence clearly the (1 + 1) dimensional boundary field theory is defined on a spatial circle $S^1$ parametrized by $\varphi$ with $t$ as the time coordinate.
The above metric possesses some useful scaling symmetries which may be described as follows
\begin{eqnarray}
r\rightarrow \lambda r,~ t\rightarrow \frac{t}{\lambda},~\varphi\rightarrow \frac{\varphi}{\lambda},~ J\rightarrow \lambda^{2} J,~ \mu\rightarrow \lambda \mu,~M\rightarrow \lambda^{2}( M-\frac {\mu^{2}}{2}\ln{\lambda}),\label{eq:sym1}\\
r\rightarrow \lambda r,~t\rightarrow \lambda t,~L\rightarrow \lambda L,~J \rightarrow \lambda J,~M\rightarrow (M-\frac{\mu^{2}}{2}\ln{\lambda}),\label{eq:sym2}
\end{eqnarray}
The symmetry given by (\ref{eq:sym1}) may be used to set $M=1$ and the symmetry corresponding to (\ref{eq:sym2}) may be used to set $L=1$, which leads to the following rescaled metric,

\begin{equation}
ds^{2}=-\frac{r^2}{f(r)}dt^{2}+\frac{dr^{2}}{r^{2}f(r)}+r^{2}\left(d\varphi-\frac{J}{2r^{2}}dt\right)^{2},\label{eq:BTZJ}
\end{equation}
where,
\[
f(r)=1-\frac{1}{r^{2}}+\frac{J^{2}}{4r^{4}}-\frac{\mu^{2}}{2r^{2}}\ln{r}.
\]

It is convenient to write down the metric (\ref{eq:BTZJ}) in the new coordinates $z=1/r$, giving
\begin{equation}
ds^2 = \frac{1}{z^{2}}\left[-f(z)dt^{2}+\frac{dz^{2}}{f(z)}+\left(d\varphi-\frac{J z^{2}}{2}dt\right)^{2}\right],\label{eq:BTz}
\end{equation}
where 
\begin{equation}
f(z)=1-z^{2}+\frac{J^{2}z^{4}}{4}+\frac{\mu^{2}z^{2}}{2}\ln{z}.\label{eq:fCR}
\end{equation}

We study our system in low angular momentum $J$ approximation. This is justified in the context of the instability of rotating black holes which is generated due to large values of angular momentum and the super radiant effect that follows from it \cite{HawkingTaylor,Berman,HawkingReall}. The low angular momentum $J$ approximation also fixes the horizon at $z=z_h=1$ which simplifies the numerical computations \cite{Nurma}. The limit $J\rightarrow 0$ is smooth as it may be seen from the metric (\ref{eq:BTz}) for an uncharged rotating BTZ black hole where the lapse function is then given by
\begin{equation}
f(z)= 1-z^{2}+\frac{J^{2}z^{4}}{4}. \label{eq:FR}
\end{equation}
The horizons of the rotating BTZ black hole are given by the zeroes of the lapse function $f(z)$ (\ref{eq:FR}) as
\begin{eqnarray}
1-z_\pm^{2}+\frac{J^{2}z_\pm^{4}}{4}=0  \nonumber \\
z^2_{+}=\frac{2}{J^2}\left[1+ \left(1-J^2\right)^{1/2}\right],\\\label{eq:zhPFR}
z^2_{-}=\frac{2}{J^2}\left[1- \left(1-J^2\right)^{1/2}\right].\label{eq:zhNFR}
\end{eqnarray}
From above equations we pick the root (\ref{eq:zhNFR}) and with the limit $J\rightarrow 0$  the expression for horizon reduces to
\begin{eqnarray}
z^2_{-} \stackrel{J\rightarrow 0}{\leadsto} \frac{2}{J^2}\left[1- \left(1-J^2+\dots \right)\right]=1+\mathcal{O}(J^2),
\label{eq:limzhFR}
\end{eqnarray}
thus in this limit only the horizon $z_{-}=1=z_h$ survives while in the same limit $J\rightarrow 0$ the other horizon $z_{+}$ goes to infinity and corresponds to a naked singularity and hence may be discarded \cite {Nurma}.

\subsection{Normal Phase}

The effect of the bulk rotation on the normal phase of the 1 + 1 dimensional boundary theory may be studied through
a charged rotating BTZ black hole with the metric given by (\ref{eq:BTz}).  The small angular momentum limit is used to fix the lapse function in this case as 
\begin{equation}
f(z)=1-z^{2}+\frac{\mu^{2}z^{2}}{2}\ln{z}. \label{eq:nrmlFCR}
\end{equation}
At this point it is worth mentioning that the bulk gauge field $A_{\alpha}$ for the non rotating case has no dependence on the angular coordinate $\varphi$ due to the diagonal nature of the metric. However for the case of the rotating charged BTZ black hole
the bulk gauge field $A_{\alpha}$ picks up a dependence on the angular coordinate $\varphi$ due to the non zero $g_{\varphi t}$ component of the metric (\ref{eq:BTz}).  Hence the anasatz for the bulk gauge field $A_{\alpha}$ for the rotating case is as follows
\begin{equation}
 A(z,\varphi) = \phi(z)dt+\xi(z) d\varphi . \label{eq:ana1}
\end{equation}
It was shown in \cite{Clement} that the rotating solution with a non-zero "angular" component for the gauge field $A_{\alpha}$ may be generated from the non rotating solution with the identification $\xi(z) = \frac{J z^{2}}{2}\phi(z)$. Thus for the small angular momentum approximation we may neglect the angular component $\xi$ of the gauge field to arrive at the following anasatz
\begin{equation}
 A(z)= A_{t}(z)dt =\phi(z)dt =\mu\ln{(\frac {z}{z_h})}dt,~~~J \ll 1,~~~\xi(z) \ll \phi(z)
\end{equation}
In the small angular momentum $J$  limit the Hawking temperature for the charged rotating BTZ black hole in the bulk is
\begin{equation}
 T_{h}= \frac{|f'(1)|}{4\pi}=\frac{4-\mu^{2}}{8\pi}
\end{equation}

\begin{figure}[htp]
\centering
\begin{minipage}[b]{0.5\linewidth}
\includegraphics[width =2.8in,height=1.8in]{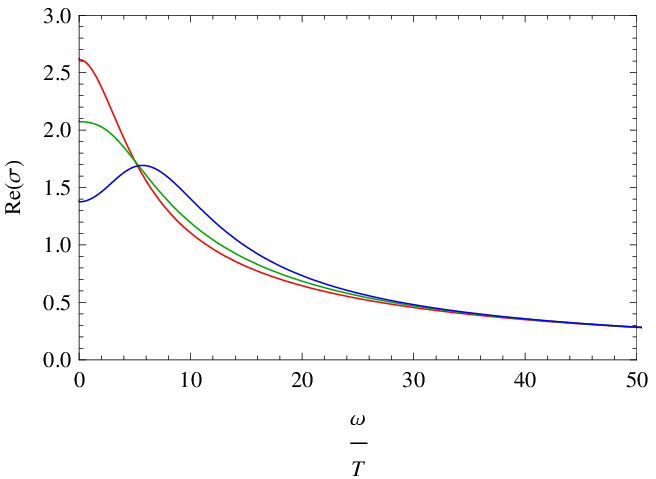}
\end{minipage}%
\begin{minipage}[b]{0.5\linewidth}
\includegraphics[width =2.8in,height=1.8in]{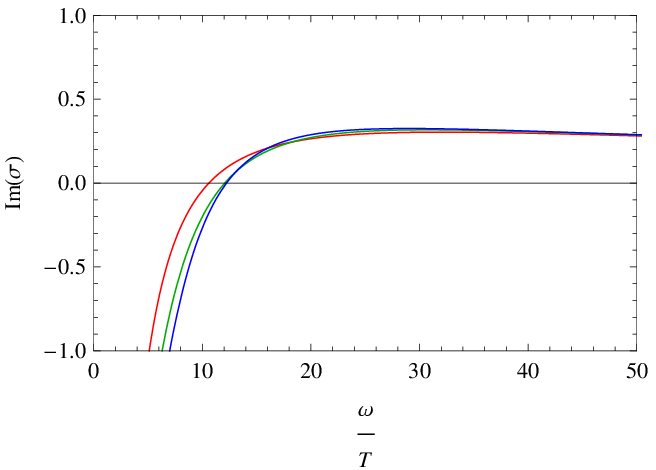}
\end{minipage}\quad
\begin{minipage}[b]{0.5\linewidth}
\includegraphics[width =2.8in,height=1.8in]{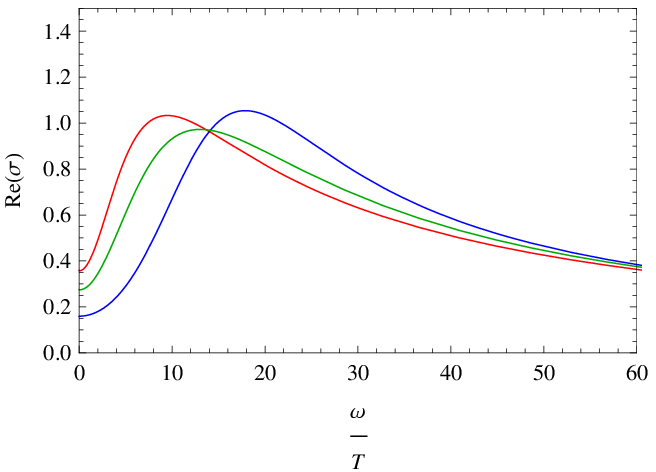}
\end{minipage}%
\begin{minipage}[b]{0.5\linewidth}
\includegraphics[width =2.8in,height=1.8in]{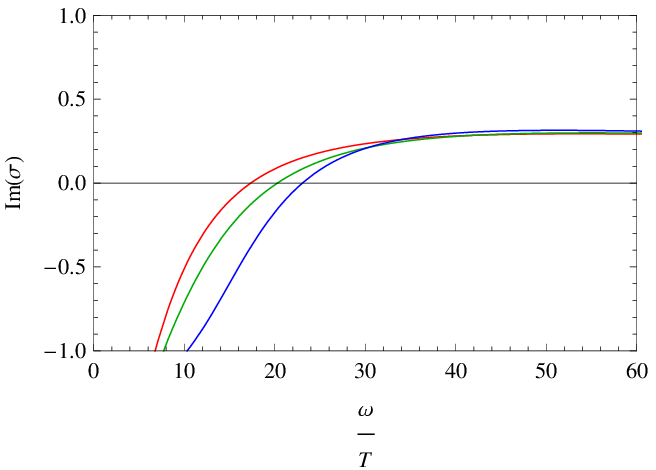}
\end{minipage}%
\caption{\label{fig:sigmaTJ} Real and imaginary parts of the conductivity for normal phase are plotted with respect to $\omega/T$. The red, green and blue curves correspond to $(J_r= 5,\mu=1.1)$ , $(J_r= 3,\mu=1.1)$ and  $(J_r= 1,\mu=1.1)$ respectively for the top two graphs while the graphs at the bottom correspond to $\mu=1.5$ for the same values of $J_r$ . } \end{figure}

To compute the ac conductivity for the normal phase of the 1+1 dimensional boundary theory we add the vector perturbation $e^{-i\omega t}A_{\varphi}$ and the metric perturbation $e^{-i\omega t}g_{\varphi t}$ to the fixed background given by the 
charged rotating BTZ black hole. The equations of motion for the bulk field are given as,
\begin{eqnarray}
R_{mn}-\frac{1}{2} g_{mn}(R+\frac{2}{L^2})=0,\label{eq:Einseq}\\
\partial_{n}(\sqrt{-g}F^{nm})=0.
\end{eqnarray} 
The perturbations for the metric and the gauge field leads to two coupled equations in terms of $g_{\varphi t}$ and $ A_{\varphi}$. Upon eliminating $g_{\varphi t}$ from the equations of motion we arrive at a linearized equation for $A_{\varphi}$, which in the small angular momentum $J$ approximation is given as 
\begin{equation}
A_{\varphi }''(z)+\left(\frac{1}{z}+\frac{ f'(z)}{f(z)}\right)A_{\varphi }'(z)+\left(\frac{\omega^{2} }{f(z)^{2}}-\frac{\mu^{2}}{f(z)}+\frac{J \omega^{2} z^{2}}{2 f(z)^{3}}\right)A_{\varphi}(z)=0. \label {eq:normaleqnAphi}
\end{equation}
Notice that although in the small angular momentum approximation the lapse function is independent of  the angular momentum $J$, the bulk equations of motion acquires a  dependence on the angular momentum $J$ from the $g_{t\varphi}$ component of the metric. This  is evident from  the last term of the above equation which is linearly dependent on the angular momentum $J$.
The form of the solution for  $A_{\varphi}$ near the boundary may be expressed as
\begin{equation}
A_{\varphi} = {\cal A}\ln(z)+{\cal B}+\cdots,\label{eq:bdryAphi}
\end{equation} 
while the near horizon  expansion of $A_{\varphi}$ with the incoming wave boundary condition is a bit subtle due to the presence of the $J$ dependent term in the equation (\ref{eq:normaleqnAphi}) which must be regularized. In accordance with \cite{Bala} we redefine the coordinate $z$ and the angular momentum $J$ in terms of the variables $ \bar {z} $
and $J_r$ as
\begin{equation}
z^{2}=z^{2}_{h}(1- \varepsilon \bar{z}^{2}) ,~~ J = \varepsilon \bar{z}^{2} J_{r},~~z_{h}-z \approx \frac{z_{h}\varepsilon\bar{z}^{2}}{2}, \label{eq:redef}
\end{equation}
where $\varepsilon$ is an infinitesimal quantity such that $z \rightarrow z_{h}$ in the limit $\varepsilon \rightarrow 0$. For the horizon boundary conditions  we assume the following ansatz for  $A_{\varphi}(z)$ near the horizon ($z_h=1$) ,
\begin{equation}
A_{\varphi}(z)= f(z)^{\delta} S(z),~~S(z)=1+ a(1-z)+ b (1-z)^2+\cdots ,\label{eq:Seqn}
\end{equation}
where the coefficients $a$ and $b$ are functions of $\mu,J_r$ and $\omega$ and the dots represent terms in higher powers of $(1-z)$.
With the substitution $A_{\varphi}(z)= f(z)^{\delta} S(z)$, the near horizon limit ($z\rightarrow 1$) of the equation (\ref{eq:normaleqnAphi}) yields the following expression for $\delta$ ,
\begin{equation}
 \left(\omega ^2+\frac{J z^2 \omega ^2}{2 f(z)}+\delta ^2 f'(z)^2 \right)\biggr\rvert_{z=1}= 0. \label{eq:smldelta}
\end{equation}
Notice that the second term in above equation is divergent and it must be regularized by using the redefinitions 
 (\ref{eq:redef}). This renders the equation (\ref{eq:smldelta}) to the following form,
\begin{eqnarray}
\delta ^2 \left(\mu ^2-4\right)^3+4 \omega ^2 \left(-2 J_r+\mu ^2-4\right)= 0,~~
\delta_{\pm}= \pm \frac{2 \omega  \sqrt{2 J_r-\mu ^2+4}}{\sqrt{\left(\mu ^2-4\right)^3}}.
\end{eqnarray}
Taking $\delta=\delta_{+}$, the near horizon form of $A_{\varphi}$ with the incoming wave boundary condition may be expressed as,
\begin{equation}
A_{\varphi}(z)|_{z \rightarrow 1} = (f(z))^{\frac{2 \omega  \sqrt{2 J_r -\mu ^2+4}}{\sqrt{\left(\mu ^2-4\right)^3}}}(1+\cdots).\label{horAphi}
\end{equation}
Now we use the equations (\ref{eq:green}) and (\ref{eq:bdryAphi}) to arrive at an expression for the  ac conductivity in terms of the value of the gauge field $A_{\varphi}$ near a small cutoff $z=\epsilon$  at the boundary as
\begin{equation}
\sigma(\omega) = \frac{i(A_{\varphi}- z A'_{\varphi} z \ln{z})}{\omega (z A'_{\varphi})}|_{z \rightarrow \epsilon}.\label{eq:AcCond}
\end{equation} 
Using the above expression in (\ref{eq:AcCond}) we numerically compute the ac conductivity for the normal phase of the 1+1 dimensional boundary theory. The real and imaginary parts of the conductivity for various values of the chemical potential $\mu$ and the redefined angular momentum $J_r$ are displayed in Fig.(\ref{fig:sigmaTJ}). \footnote {All the graphs in this article have been computed with modified numerical codes from those of  S. Hartnoll and C. P. Hertzog generously provided on their website.}

We observe from Fig.(\ref{fig:sigmaTJ}) that the profiles for both the real and the imaginary parts of the ac conductivity 
as a function of $\omega$ are similar to those of the boundary field theory dual to a charged non rotating BTZ black hole  \cite {Ren}. However unlike the non rotating case the profiles depend on the angular momentum $J_r$ of the rotating black hole. 
It is seen from Fig. (1)  that the peak of the individual graphs for the real part of ac conductivity increases for higher values of the angular momentum $J_r$ with a fixed chemical potential  $\mu$. Furthermore the peak of the conductivity curves shift towards the higher frequency side with increasing values of the chemical potential $\mu$ which is also a behaviour similar to the case of the boundary field theory discussed in \cite {Ren}.  The real part of the ac conductivity still has a delta function at $\omega=0$ just as in the case for the non-rotating BTZ black hole in the bulk described in subsections 3.1 and 3.2.

From the perspective of the 2+1 dimensional bulk AdS space time the difference obviously arises from the angular momentum $J$ dependent term in the equation for the gauge field perturbation $A_{\varphi}(z)$ in (\ref {eq:normaleqnAphi}) in contrast to the corresponding equation for the non rotating case described in \cite {Ren}. Comparing the two equations (\ref{eq:normaleqnAphi}) and (\ref{eq:pertAx})  we observe that the additional term for the rotating case under consideration, constitutes an extra contribution to the expression for the $\omega^2$ dependent part in (\ref {eq:normaleqnAphi}). Given that the conductivity for the boundary theory is obtained from the gauge field correlators in the bulk , the modification of the $\omega^2$ dependent part in  (\ref {eq:normaleqnAphi}) is the origin for the observed dependence of conductivity on $J$ or the factor $J_r$.

 As shown in \cite {Bala}, the 1+1 dimensional boundary field theory following from a BTZ black hole in the bulk, is expected to be in some strongly coupled phase of a Fermi-Luttinger liquid \cite {Luttinger, Mattis, Voit, Amit}. This was the case with both the uncharged rotating Chern Simmons- BTZ black hole with bulk fermions and Wilson lines \cite {Bala} and the charged  BTZ black hole \cite {Nilanjan, Ren}. For the rotating charged BTZ black hole considered by us notice that $\varphi$ is an angular coordinate with $( 0 \leq  \varphi  \leq  2\pi )$  in contrast to the non rotating cases. This indicates that the dual 1+1 dimensional boundary field theory is defined on a spatial circle $S^1$. Such a boundary field theory is expected to describe  some strongly coupled phase of a Fermi- Luttinger liquid on a circle in a bosonized formulation, which may be physically realized in mesoscopic systems with ring geometries \cite{ Weinmann, Wade, Haussler}. 
 
 In \cite {Bala} it has been argued in the context of a rotating Chern Simons BTZ black hole in presence of fermions that the mass of the bulk field controls the scaling dimension of the dual operator. It is further argued that the transition from the non interacting free fermion $(g=0)$ to the interacting case may be represented in the bosonized version by a scalar boson on a circle with a modified radius $R=\frac {1 + g/2 \pi}{1-g/ 2\pi}R_{free}$ which is related to the conformal dimensions of the interacting fermions. Hence this establishes a direct relation between the mass of the bulk fermions and the coupling constant $g$ of the Fermi Luttinger liquid on a circle. Notice that this is also true in our case for the rotating charged BTZ black hole in the bulk as the boundary field theory is defined on  a circle and is expected to describe some strongly coupled phase of  a Fermi-Luttinger liquid on a circle. As we will show later in the next section, for the superconducting phase of the boundary theory the angular 
momentum $J$ in the low angular momentum approximation and the scalar field winding $\alpha$ defined later, modifies the mass of the bulk charged scalar field which  is related to the strength of the coupling constant $g$ of the Luttinger liquid. Hence we would expect both the angular momentum $J$ and the winding number $\alpha$  to be related through the {\it effective mass} of  the scalar field to the coupling constant $g$ of the Fermi Luttinger liquid on a circle.


\subsection{Superconducting Phase}

The superconducting phase of the 1+1 dimensional boundary theory is dual to a bulk charged rotating BTZ black hole with
a scalar field charged under the bulk $U(1)$ gauge field. Working in the probe limit and the small angular momentum  $J$ approximation, the lapse function in the metric (\ref{eq:BTZCR}) is given by $f(z)=1-z^2$ \cite{Ren,Nurma}.
The bulk equations of motion once again are given by (\ref {eq:bulkeom}). Following the discussion in the subsection 4.1 for the normal phase of the 1+1 dimensional boundary theory we observe that both the bulk scalar field $\Psi$ and the gauge field $A_{\alpha}$ will pick up a dependence on the angular coordinate $\varphi$ for the rotating case, leading to the following ansatz 
\begin{equation}
 A(z,\varphi) = \phi(z)dt+\xi(z) d\varphi,~~~~\Psi = \psi(z,\varphi) \label{eq:supana1}.
\end{equation}
which reduces to the following under the small angular momentum ($J$) approximation as mentioned earlier,
\begin{equation}
 A(z)=A_t(z) dt = \phi(z)dt,~~~~\Psi = \psi(z,\varphi) \label{eq:supana}.
\end{equation}
The equation of motion for the scalar field may now be solved by a separation of variables through the definition $\Psi(z,\varphi)= \psi(z)S(\varphi)$ where, $S(\varphi)= \exp(i\alpha\varphi)$ \footnote{In general the parameter $\alpha$ can take complex and unconstrained values but we restrict ourselves to $\alpha~\epsilon~R$. Values of $\alpha $ should not be confused with the eigenvalues of the equation of motion for $S(\varphi)$ which is $\lambda$.}. Substituting the above expression for $\Psi$ back into the equation of motion for the scalar field we arrive at the following set of equations,
\begin{eqnarray}
\partial_{z}(\sqrt{-g}g^{zz}\partial_{z} \Psi)+D_{\varphi}(\sqrt{-g}g^{\varphi\varphi}D_{\varphi} \Psi)-i\sqrt{-g}A_{t} g^{t\varphi}\partial_{\varphi}\Psi\nonumber\\
-\sqrt{-g}\left( m^{2}+A_{t} g^{tt}A_{t}+2A_{t} g^{t\varphi}A_{\varphi}\right)\Psi = 0,\label{eq:scalar1}\\\nonumber\\
\sqrt{-g}g^{\varphi\varphi}\partial^{2}_{\varphi}S(\varphi)-i\sqrt{-g}A_{t} g^{t\varphi}\partial_{\varphi}S(\varphi)-2i\sqrt{-g}A_{\varphi}g^{\varphi\varphi}\partial_{\varphi}S(\varphi)=-\lambda S(\varphi),\label{eq:angSeq}\\\nonumber\\
\partial_{z}(\sqrt{-g}g^{zz}\partial_{z} \psi(z))-\sqrt{-g}A_{t} g^{tt} A_{t}\psi(z)-\sqrt{-g}m^{2}\psi(z) = \lambda \psi(z),\label{eq:psieq}
\end{eqnarray}

where,
\begin{equation}
\lambda = \frac{\alpha^{2}}{z}\left(1-\frac{J^{2}z^{4}}{4f(z)}\right)+ \frac{\alpha Jz}{2f(z)}A_{t}(z)-\frac{2\alpha}{z}\left(1-\frac{J^{2}z^{4}}{4f(z)}\right)A_{\phi}(z). \label{eq:lambCR}
\end{equation}
In the context of the small angular momentum approximation described in the previous section and the ansatz (\ref{eq:supana}) we set $A_{t}(z) = \phi(z)$ and $A_{\phi} = \xi(z) =0 $ in the equation  (\ref{eq:lambCR}) and neglect the term proportional to $J^{2}$ to arrive at

\begin{equation}
\lambda = \frac{\alpha^{2}}{z}+ \frac{\alpha Jz}{2f(z)}\phi(z)
\end{equation}
thus the equation (\ref{eq:psieq}) becomes 
\begin{equation}
\psi^{''}(z)+\left(\frac{f^{'}(z)}{f(z)}-\frac{1}{z}\right)\psi^{'}(z)+\left(\frac{\phi(z)^{2}}{f(z)^{2}}-\frac{m^{2}}{z^{2} f(z)}-\frac{1}{f(z)}\left[\alpha^{2}+\frac{\alpha J z^{2}}{2f(z)}\phi(z) \right]\right)\psi(z)= 0. \label{eq:psieqapp}
\end{equation}
From above equation we observe that the effective mass of the scalar field $\psi$ for the rotating case is modified from the non rotating case (16) to the following
\begin{equation}
 m^{2}_{eff}=m^2+z^2 \alpha^{2}-\frac{z^2 \phi(z)^{2}}{f(z)}+\frac{\alpha J z^{4}}{2f(z)}\phi(z). \label{eq:meff}
\end{equation}

\noindent Unlike the non rotating case the expression for the effective mass of the scalar field $\psi$ includes a positive term proportional to the factor $\alpha$ and a negative term proportional to $\phi(z)$. In the charged case where the factor $\alpha$ is zero, the $\phi(z)$ dependent term tends to decrease the effective mass of the scalar field making the condensation easier \cite{Gubser}. However for the rotating case when the factor $\alpha$ is non zero the terms in $\alpha$ tend to increase the effective mass of the scalar field making the condensation harder. Thus a competitive behaviour exists between the $\alpha$ dependent term and $\phi(z)$ dependent term in the expression for the effective mass of the scalar field. We will discuss the effect of $\alpha$ on the formation of condensate through both analytical and numerical studies in the next section.

The three components of the Maxwell equation for the bulk gauge field may be expressed as
\begin{eqnarray}
\partial_{m}(\sqrt{-g}F^{m z})+i\sqrt{-g}\left(\Psi \partial^{z} \Psi^{*}-\Psi^{*}\partial^{z} \Psi \right)= 0,\label{eq:MxEom1}\\
\partial_{m}(\sqrt{-g}F^{m \varphi})+i\sqrt{-g}\left(\Psi (D^{\varphi} \Psi)^{*}-\Psi^{*}D^{\varphi} \Psi \right)= 0,\label{eq:MxEom2}\\
\partial_{m}(\sqrt{-g}F^{m t})+i\sqrt{-g}\left(\Psi (D^{t} \Psi)^{*}-\Psi^{*}D^{t} \Psi \right)= 0 .\label{eq:MxEom3}
\end{eqnarray}
From above we observe that the equations (\ref{eq:MxEom2}) and (\ref{eq:MxEom3}) reduce to
\begin{eqnarray}
\partial_{z}(\sqrt{-g}g^{\varphi t}g^{zz}F_{zt})-2\sqrt{-g}\Psi^{2} g^{\varphi t} A_{t} = 0,\nonumber\\
\partial_{z}(\sqrt{-g}g^{zz}g^{tt}F_{zt})-2\sqrt{-g}\Psi^2 g^{tt}A_{t} = 0.\label{eq:Mxeom}
\end{eqnarray}
The equations in (\ref{eq:Mxeom}) may be combined together to provide a single equation of motion for the scalar part of  the gauge field \footnote{In \cite{Nurma} the term proportional to $J$ in the equation of motion for the scalar part $\phi(z)$ of the gauge field is erroneously computed to be equal to $J z^3$ .} in the small angular momentum approximation as 
\begin{equation}
\phi^{''}(z)+\frac{1}{z} \left(1+J z^{2}\right)\phi^{'}(z)-\frac{2 \psi(z)^{2}}{z^{2} f(z)}\phi(z)= 0.\label{eq:phieomJ}
\end{equation} 

We observe that in the limit $J \rightarrow 0$, equations (\ref{eq:phieomJ}) and (\ref{eq:psieqapp}) reduce to that for the case of the charged non-rotating BTZ black hole as described in \cite{Ren,Nurma}.

\subsection{Analytical Solution for the Condensate}

To investigate the instability of the scalar field that leads to formation of scalar hair for the charged rotating BTZ black hole in the bulk we consider the mass of the  scalar field to be near the BF bound i.e. $m^{2}=m^{2}_{BF}=-1$ \cite{Breitenlohner}. Recall that this bulk instability translates to the superconducting phase transition and the formation of a charged condensate in the 1+1 dimensional boundary field theory.  For this we need to consider the boundary conditions for the gauge field $\phi(z)$ and scalar field $\psi(z)$ near the horizon $z_{h}=1$ as,
\begin{equation}
\phi(z_{h})=0,~~~~\psi^{'}(z_{h})=-\frac{(m^{2}+\alpha^{2}z^{2}_{h})}{2 z_{h}}\psi(z_{h})= \frac{(1-\alpha^{2}z^{2}_{h})}{2 z_{h}}\psi(z_{h}).\label{eq:BCJ}
\end{equation}
In the small angular momentum approximation and in the probe limit we may assume that the lapse function is given as
\begin{equation}
 f(z)= 1- \frac{z^{2}}{z^{2}_{h}}
\end{equation}
Following \cite{GregorySoda} and using the boundary conditions mentioned above we may express the near horizon ($z=z_{h}=1$) expansions of  the fields $\phi(z)$ and $\psi(z)$, up to second order as
\begin{eqnarray}
\phi(z) = \phi^{'}(z_{h})(z-z_{h})+\frac{1}{2}\phi^{''}(z_{h})(z-z_{h})^{2}+\cdots, \label{eq:phihexp} \\
\psi(z) = \psi(z_{h})+\psi^{'}(z_{h})(z-z_{h})+\frac{1}{2}\psi^{''}(z_{h})(z-z_{h})^{2}+\cdots \ \ . \label{eq:psihexp} 
\end{eqnarray}
Using the  redefinition of the coordinate $z$ and the angular momentum $J$ in terms of the variables $ \bar {z} $
and $J_r$ as given by equation (\ref{eq:redef}), we compute the coefficients of the second order terms in the near horizon expansions for the fields $\phi(z)$ and $\psi(z)$. With the help of the $\phi(z)$ equation of motion (\ref{eq:phieomJ}) we arrive at an expression for $\phi^{''}(z)$ at $ z=z_h$ as
\begin{equation}
\phi^{''}(z_{h})+\left(1+\varepsilon \bar{z}^{2} z^{2}_{h} J_{r}\right)\frac{\phi^{'}(z_{h})}{z_{h}}+ \frac{\psi^{2}(z_{h})\phi^{'}(z_{h})}{z_{h}}=0.\label{eq:phidpeom}
\end{equation}
In the limit $\varepsilon \rightarrow 0$ equation (\ref{eq:phidpeom}) leads to
\begin{equation}
\phi^{''}(z_{h})= - \frac{1}{z_{h}}\left(1+\psi(z_{h})^{2}\right)\phi^{'}(z_{h}).\label{eq:phidpzh}
\end{equation}
Hence using (\ref{eq:phidpzh}) we may write the modified near horizon expansion for $\phi(z)$ as,
\begin{equation}
\phi(z) = \phi^{'}(z_{h})(z-z_{h}) - \frac{1}{2 z_{h}}\left(1+\psi(z_{h})^{2}\right)\phi^{'}(z_{h})(z-z_{h})^{2}+\cdots \ \ .\label{eq:phiJexp} 
\end{equation}
Now from the equation of motion of $\psi(z)$  in ( \ref {eq:psieqapp}), we observe that the term proportional to $J$ is given as
\[ -\frac{\alpha Jz^{2}}{2 f^{2}}\phi(z) \psi(z) \] ,
which is divergent at the horizon $z=z_h$. The expression for this divergent term in the near horizon $(z=z_h=1)$ limit is given by
\begin{eqnarray}
\left.-\frac{\alpha Jz^{2}}{2 f^{2}}\phi(z)\psi(z)\right|_{z=z_{h}}= -\frac{\alpha J_{r}z^{3}_{h}}{4}\phi^{'}(z_{h})\psi(z_{h})
+ \nonumber \\ \frac{\varepsilon \bar{z}^{2} J_{r}\alpha z^{4}_{h}}{2}\left(\frac{\psi(z_{h})\phi^{''}(z_{h})}{8 z^{2}_{h}}+\frac{\psi^{'}(z_{h})\phi^{'}(z_{h})}{4 z^{2}_{h}} \right),
\end{eqnarray} 
hence in the limit $\varepsilon\rightarrow 0$ we have
\begin{eqnarray}
 \left.-\frac{\alpha Jz^{2}}{2 f^{2}}\phi(z)\psi(z)\right|_{\varepsilon \rightarrow 0}= -\frac{\alpha J_{r}z^{3}_{h}}{4}\phi^{'}(z_{h})\psi(z_{h}). \label {eq:regJ}
\end{eqnarray}

Using the equation (\ref{eq:regJ}) the equation of motion for the charged scalar field (\ref{eq:psieqapp}) in the near horizon limit becomes
\begin{eqnarray}
\psi^{''}(z_{h})+\left.\left(\frac{-(z^3+z z^2_{h})\psi^{'}(z)+z^2_{h}(1-\alpha^2 z^2)\psi(z)}{z^2(z^2-z^2_{h})}\right)\right|_{z=z_{h}}\nonumber \\+\left.\left(\frac{\phi(z)^{2}\psi(z)}{f(z)^{2}}-\frac{\alpha Jz^{2}\phi(z)\psi(z)}{2 f^{2}(z)}\right)\right|_{z=z_{h}}= 0.
\end{eqnarray}
The above equation may be further reduced to obtain an expression for $\psi^{''}(z)$ at $z=z_h$ as
\begin{eqnarray}
\psi^{''}(z_{h})= -\frac{1}{8 z^2_{h}}\left(3+2\alpha^2 z^2_{h}-\alpha^4 z^4_{h}+z^4_{h}\phi^{'}(z_{h})^2+\alpha J_{r}z^5_{h}\phi^{'}(z_{h})\right)\psi(z_{h}).\label{eq:psipp}
\end{eqnarray}
Using equations (\ref{eq:BCJ}),(\ref{eq:psihexp}) and (\ref{eq:psipp}) we may write the modified near horizon expansion for the scalar field  $\psi(z)$ as
\begin{eqnarray}
\psi(z)= \psi(z_{h})+\frac{\psi(z_{h})(1-\alpha^{2}z^{2}_{h})}{2 z_{h}}(z-z_{h})
 -\frac{\psi(z_{h})}{16 z^2_{h}}\nonumber \\\left(3+\alpha^2 z^2_{h}(2-\alpha^2 z^2_{h})+z^4_{h}\phi^{'}(z_{h})^2+\alpha J_{r}z^5_{h}\phi^{'}(z_{h})\right)(z-z_{h})^2 +\cdots \ \ \ .\label{eq:psiJexp}
\end{eqnarray}

Near the boundary $(z \rightarrow 0)$ the bulk gauge field $ \phi(z)$ and the scalar field $\psi (z)$ are expressed as
\begin{eqnarray}
\phi(z) = \mu \ln{z}- \rho, ~~~ \psi(z)= \psi_{1} z\ln{z}+\psi_{2} z.
\end{eqnarray}

The scaling dimensions $\Delta_{-}$ and $\Delta_{+}$ of the corresponding dual operators $\psi_{1}$ and $\psi_{2}$ are related to the mass of bulk scalar field $\psi$ through the relation $\Delta_{\pm}=1 \pm \sqrt{m^2+1}$. For $-1\geq m^2\leq 0$, we can choose any of the $\psi_{1,2}$ to condense. For $m^2\geq 0$, we can only take $\psi_2$ to condense. For our case we take the mass of the bulk scalar field near the BF bound ($m^2= m^2_{BF}= -1$) which makes the condensation of $\psi_{1,2}$ possible. Now due to rotation and nonzero bulk gauge field $\phi$, the mass $m^2$ of the bulk scalar field $\psi$ gets modified to $ m^2_{eff}$ which may be observed through the equation (\ref{eq:meff}). Here for a non zero value of $\alpha$  the effective mass of the bulk scalar field is pushed away from the BF bound making the condensation of $\psi_{1,2}$ harder while a nonzero bulk gauge field $\phi$ pushes the effective mass towards the BF bound making the condensation easier. Thus we may state that the parameter $\alpha$ plays a vital role in formation of the condensates $\psi_{1,2}$ by affecting the effective mass of the bulk scalar field $\psi$.

We take $\psi_{1}=0$, in order to study the condensation of the operator ${\cal O}_2$ dual to $\psi_{2}$. We begin with sewing the horizon and the boundary expansions of the fields $\phi(z)$ and $\psi(z)$ near $z= z_{h}/2$. We also match the derivatives of the boundary and the horizon expansions for the fields near the sewing point. This results in the following set of equations
\begin{eqnarray}
\frac{z_{h}\psi_{2}}{2}=\psi(z_{h})-\frac{\psi(z_{h})(1-\alpha^{2}z^{2}_{h})}{4}\nonumber\\
 -\frac{\psi(z_{h})}{64}\left(3+\alpha^2 z^2_{h}(2-\alpha^2 z^2_{h})+z^4_{h}\phi^{'}(z_{h})^2+\alpha J_{r}z^5_{h}\phi^{'}(z_{h})\right),\label{eq:sew1}
\\
\psi_{2} = \frac{\psi(z_{h})(1-\alpha^{2}z^{2}_{h})}{2 z_{h}}\nonumber\\
 -\frac{\psi(z_{h})}{32 z_{h}}\left(3+\alpha^2 z^2_{h}(2-\alpha^2 z^2_{h})+z^4_{h}\phi^{'}(z_{h})^2+\alpha J_{r}z^5_{h}\phi^{'}(z_{h})\right),\label{eq:sew2}
\\
 \mu\ln(\frac{z_H}{2})-\rho = -\frac{z_{h}}{2}\phi^{'}(z_{h})-\frac{z_{h}}{8}\phi^{'}(z_{h})\left(1+L^2\psi^2(z_{h})\right),\label{eq:sew3}
\\
\frac{2\mu}{z_{h}}= \phi^{'}(z_{h})+\frac{1}{2}\left(1+\psi^2(z_{h})\right)\phi^{'}(z_{h}).\label{eq:sew4}
\end{eqnarray}
From equations (\ref{eq:sew3}) and (\ref{eq:sew4}) we arrive at the following relations
\begin{equation}
\phi^{'}(z_{h})=-4 a \mu ,~~~ \psi(z_{h})^2=-\frac{1}{a}(1+3 a), \label{eq:sewphipsi}
\end{equation}
where, $a=\ln{\frac{1}{2}}+\frac{1}{2}$ and as a  consequence of the boundary condition $\phi(z_{h})=0$ at the horizon we have $\rho=\mu\ln{(z_{h})}$. Similarly from the equations (\ref{eq:sew1}) and (\ref{eq:sew2}) we have

\begin{eqnarray}
\phi^{'}(z_h)&=&\frac{\sqrt{3} \sqrt{3 \alpha ^2 J_{r}^2 z_h^{10}+12 \alpha ^4 z_h^8+104 \alpha ^2 z_h^6+92 z_h^4}-3 \alpha  J_r z_h^5}{6 z_h^4},\label{eq:sewphi1}\\
\psi_{2}&=& \frac{7}{6 z_{h}}\psi(z_{h}).\label{eq:condpsi}
\end{eqnarray}
Assuming  $J_r$ and $\alpha$ small we may expand the equation (\ref{eq:sewphi1}) in $\alpha$ and $J_r$ and retain only the quadratic order terms to arrive at the following expression for $\phi^{'}(z_h)$ 
\begin{eqnarray}
\phi^{'}(z_{h})= \frac{- J_{r}z_{h} \alpha}{2}+ \frac{1}{z^2_{h}}(1+\frac{13}{23}\alpha^2 z^2_{h}).\label{eq:sewphi}
\end{eqnarray}

\begin{figure}[htp]
\centering
\begin{minipage}[b]{0.5\textwidth}
\includegraphics[width =2.8in,height=1.8in]{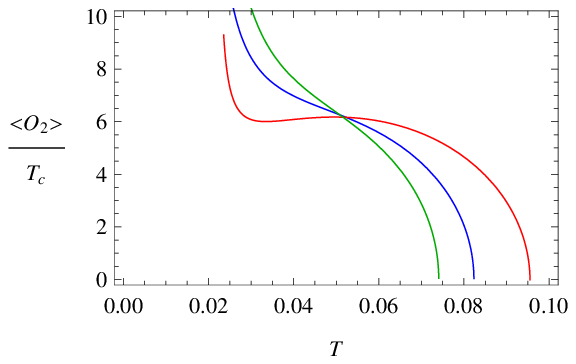}
\end{minipage}%
\begin{minipage}[b]{0.5\textwidth}
\includegraphics[width =2.8in,height=1.8in]{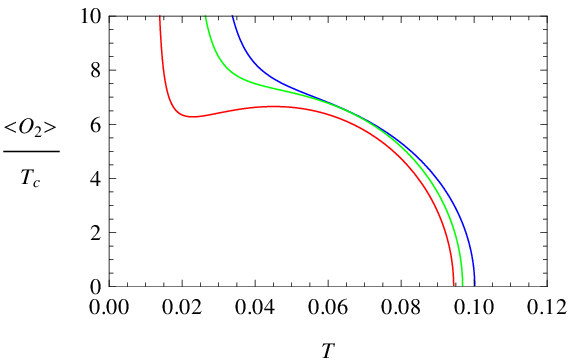}
\end{minipage}%
\caption{\label{fig:CondensTHJ}The above graphs show the theoretical plot for  $<{\cal O}_{2}>/T_{c}$  vs  $T$. The graph on the left side is for different values of $\alpha$ with fixed $J_{r}=0.1$ where, $(0.1,0.3)$ is the blue curve,  $(0.1,0.4)$ is the green curve and $(0.1,0.03)$ is the red curve. The graph on the right side is for different values of $J_{r}$ for fixed $\alpha=0.03$ with, $(1,0.03)$ as the blue curve,  $(0.5,0.03)$ as the green curve and $(0.1,0.03)$ as the red curve .}
\end{figure}

Now using the equations (\ref{eq:sewphipsi}), (\ref{eq:condpsi}) and (\ref{eq:sewphi}), the expression for the expectation value\footnote{Our result for the expectation value of $\cal O$ varies from the result of \cite{Nurma} because we completely eliminate the dependence on the factor $a$ between the expressions of $\phi^{'}(z_{h})$ and $\psi(z_{h})$ as given in (\ref{eq:sewphipsi}).} 
of operator ${\cal O}_{2}$ dual to $\psi_{2}$ may be expressed as
\begin{eqnarray}
\left<{\cal O}_{2}\right> \approx \frac{7\pi T }{3}\sqrt{\frac{\mu+\sqrt{\frac{3}{23}}\left(-\frac{23 \pi  T}{2}+\frac{\sqrt{69} J_{r} \alpha }{32 \pi ^2 T^2}-\frac{13 \alpha ^2}{8 \pi T }\right)}{-\frac{J_{r} \alpha }{32 \pi^2 T^2}+\frac{\pi}{2}\sqrt{\frac{23}{3}}\left(T+\frac{13 \alpha^2}{92 \pi^2 T}\right)}},\label{eq:condenO2}
\end{eqnarray}
where, $T = 1/2\pi z_{h} $  .We plot $\left<{\cal O}_{2}\right>/T_{c}$ from the analytical expression in (87), 
for different sets of values for the parameters $\alpha$ and $J_{r}$ as shown in Fig.(\ref{fig:CondensTHJ}). The expression for $T_c$ may be obtained from equating numerator in the expression of $\left<{\cal O}_{2}\right>$ to zero.
\begin{figure}[htp]
\centering
\begin{minipage}[b]{0.5\linewidth}
\includegraphics[width =2.8in,height=1.8in]{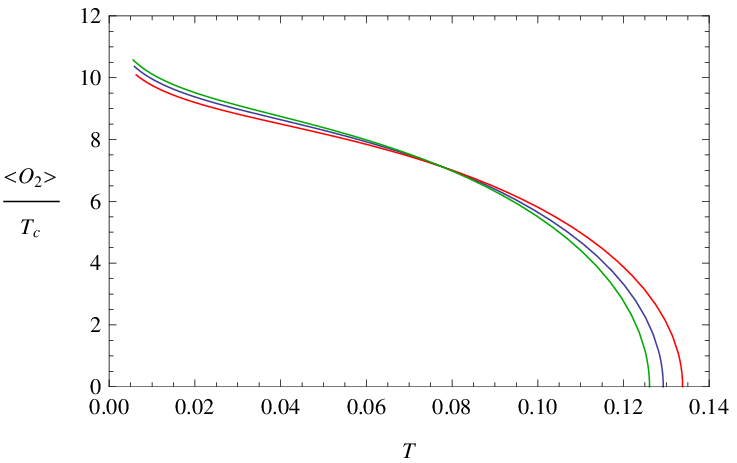}
\end{minipage}%
\begin{minipage}[b]{0.5\linewidth}
\includegraphics[width =2.8in,height=1.8in]{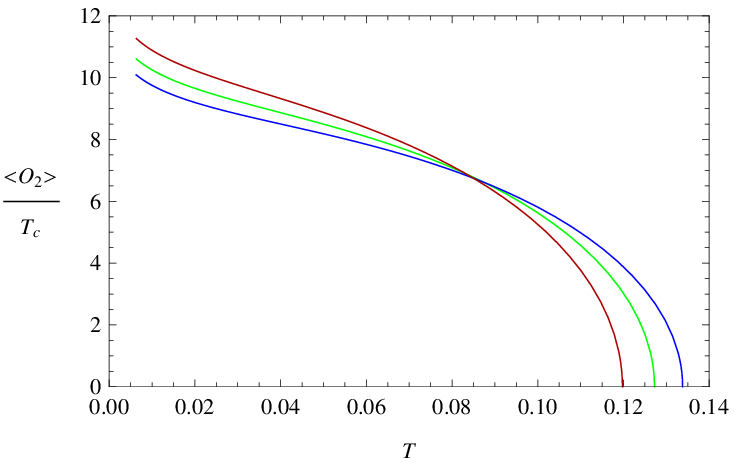}
\end{minipage}%
\caption{\label{fig:CondensNUMJ}The above graphs show the numerical plot for  $<{\cal O}_{2}>/T_{c}$  vs  $T$. The graph on the left side is for different values of $\alpha$ with fixed $J_{r}=0.1$ where, $(0.1,0.3)$ is the blue curve,  $(0.1,0.4)$ is the green curve and $(0.1,0.03)$ is the red curve. The graph on the right side is for different values of $J_{r}$ for fixed $\alpha=0.03$ with, $(1,0.03)$ as the blue curve,  $(0.5,0.03)$ as the green curve and $(0.1,0.03)$ as the red curve .}
\end{figure}

From the plots in Fig.(\ref{fig:CondensTHJ}) we observe that for fixed value of $J_{r}$ the critical temperature decreases with increasing values of  $\alpha$ which makes the formation of the condensate harder.  This is entirely expected from the discussions following (60). Whereas for fixed value of $\alpha$ the critical temperature increases with
increasing values of $J_{r}$ as may be seen from the analytical expression (87). From  the expression for $\left<{\cal O}_{2}\right>$ in (\ref {eq:condenO2}) we also observe that
it vanishes at the zeroes of the numerator inside the square root.  This leads us to a critical value $J_{r}=J^{c}_r$  in terms of
 $\mu$ and the temperature $T $ as 
\begin{equation}
J^{c}_r = 16 \sqrt{\frac{23}{3}} \pi^3 \frac{T^3}{\alpha} + \frac{52 \pi  T \alpha}{\sqrt{69}} - \frac{32 \pi^2 T^2 \mu}{3\alpha}.\label{eq:Jcritical}
\end{equation}
For fixed values of the chemical potential $\mu$,  the temperature $T $ and the scalar field winding $\alpha$,  we observe that the critical value $J^{c}_r$ establishes a lower bound on the angular momentum of the rotating black hole in the bulk for the superconducting phase transition to occur. This conclusion is in line with the relation between the angular momentum $J$ and the coupling constant $g$ of the Fermi-Luttinger liquid which is expected to describe the boundary field theory. In which case the critical value $J^c_r$ for a fixed chemical potential $\mu$ and a temperature $T$  then indicates a critical value  of the coupling constant $g=g_c$ for the superconducting phase transition in the boundary Fermi-Luttinger liquid. Notice that a similar exercise may be done to obtain a critical value for the scalar field winding $\alpha$ also. In fact as we have argued earlier both $J$ and $\alpha$ modifies the effective mass of the scalar field and are hence both related to the coupling constant $g$ of the boundary Fermi-Luttinger liquid.

We proceed further with calculating the condensate $<{\cal O}_{2}>$ numerically for different values of $J_{r}$ and the parameter $\alpha$ as shown in Fig.(\ref{fig:CondensNUMJ}). It may be observed from these numerical plots that for the condensate $<{\cal O}_{2}>$ the critical temperature decreases with increasing values of $\alpha$ for a fixed value of $J_{r}=0.1$ and it increases for increasing values of $J_r$ for a fixed value of $\alpha=0.03$. 

From the Figs. (\ref{fig:CondensTHJ}) and (\ref{fig:CondensNUMJ}) the temperature dependence of the condensate and its variation with the angular momentum $J_r$ and the parameter $\alpha$ are broadly similar for the analytical and the numerical plots. For  varying $\alpha$ and $J_r$ ( the first graphs in Fig.(\ref{fig:CondensTHJ}) and Fig.(\ref{fig:CondensNUMJ}) ) the difference in  the actual condensate profiles are within acceptable range with the critical temperature $T_c$ ranging between $(0.076-0.098)$ for the analytical plots and between $(0.127-0.135) $ for the numerical plots. This maybe attributed to the different approximations used for the analytical and the numerical computation as per the standard practice for obtaining a stable numerical solution \cite {Hartnoll1,Hartnoll2,Ren}. From this it may be noted that the numerical solution is possibly more accurate than the corresponding analytical solution where only quadratic order terms have been retained in the expression for $\phi^{'}(z_h)$ in the equation (\ref{eq:sewphi1}).

Comparing with the non rotating charged black hole \cite {Ren} we notice that although the condensate profiles in our case are similar the dependence on the angular momentum $J_r$ and the scalar field winding $\alpha$ around the compact $\varphi$ direction, illustrates the role of these parameters on the condensate formation and the critical temperature $T_c$ in the superconducting phase of the boundary field theory. Recall from the earlier arguments in Subsection (4.1) that both the angular momentum $J_r$ and the scalar field winding $\alpha$ are related to the coupling constant $g$ of the Fermi Luttinger liquid which is expected to describe the boundary field theory on a circle. So our findings from both the analytical and the numerical treatments are in conformity with this expectation.

\subsection{Conductivity for the Superconducting Phase}

To obtain the ac conductivity for the superconducting phase we follow the now standard procedure of adding vector perturbation $e^{-i\omega t}A_{\varphi}$ to the fixed bulk background. From the Maxwell equations (\ref{eq:MxEom2}) and (\ref{eq:MxEom3}) we arrive at the linearized equation for $A_{\varphi}$, which in the small angular momentum approximation may be written as 

\begin{equation}
A_{\varphi }''(z)+\left(\frac{1}{z}+J z +\frac{ f'(z)}{f(z)}\right)A_{\varphi }'(z)+\left(\frac{\omega^{2} }{f(z)^{2}}+\frac{J \omega^{2} z^{2}}{2 f(z)^{3}}-\frac{2 \psi(z)^{2}}{z^2 f(z) }\right)A_{\varphi}(z)=0.\label{eq:supeqAphi}
\end{equation}  
An analytic solution for the above differential equation seems computationally intractable,  hence we solve it numerically.
The form of the solution for  $A_{\varphi}$ near the $AdS_3$ boundary may be written as
\begin{equation}
A_{\varphi} = {\cal A}\ln(z)+{\cal B}+\cdots
\end{equation} 
\begin{figure}[H]
\centering
\begin{minipage}[c]{0.5\linewidth}
\includegraphics[width =2.8in,height=1.8in]{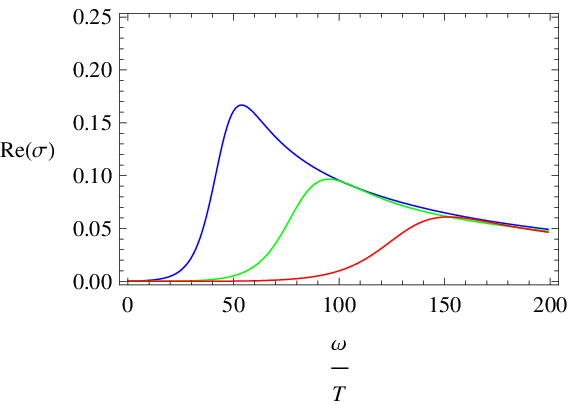}
\end{minipage}%
\begin{minipage}[c]{0.5\linewidth}
\includegraphics[width =2.8in,height=1.8in]{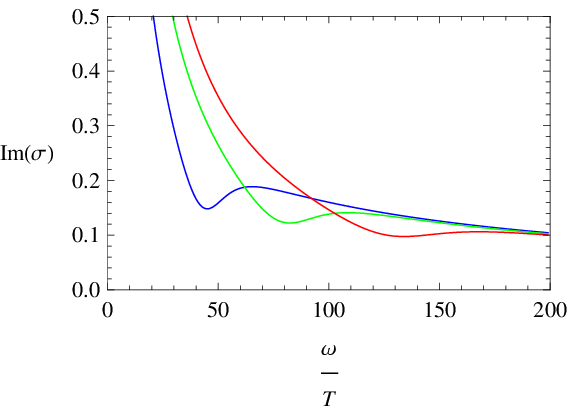}
\end{minipage}\quad
\begin{minipage}[c]{0.5\linewidth}
\includegraphics[width =2.8in,height=1.8in]{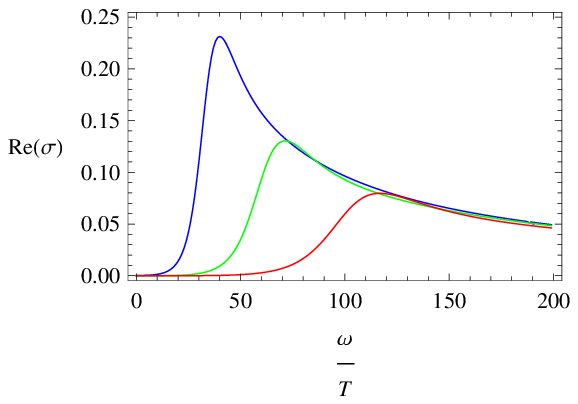}
\end{minipage}%
\begin{minipage}[c]{0.5\linewidth}
\includegraphics[width =2.8in,height=1.8in]{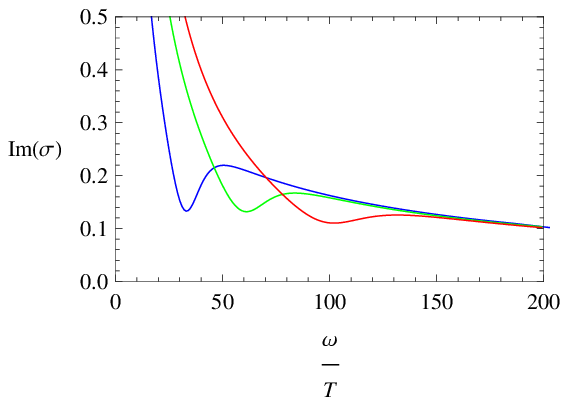}
\end{minipage}%
\caption{\label{fig:scalarcond1}Real and imaginary parts of the ac conductivity for the condensate ${\cal O}_{1}$ are plotted with respect to $\omega/T$ for different values of $J_r$ and $\alpha$. The top two graphs are  for ($J_r=2,~\alpha=0.001$) where the blue, green and the red curves correspond to different values of the temperature $T/T_c=0.612, 0.337, 0.208$ respectively for both the real and the imaginary part of  the ac conductivity. The bottom two graphs are for ($J_r=0.02,~\alpha=0.001$) with the same values of the temperature.} 
\end{figure}

\begin{figure} [H]
\centering
\begin{minipage}[c]{0.5\linewidth}
\includegraphics[width =2.8in,height=1.8in]{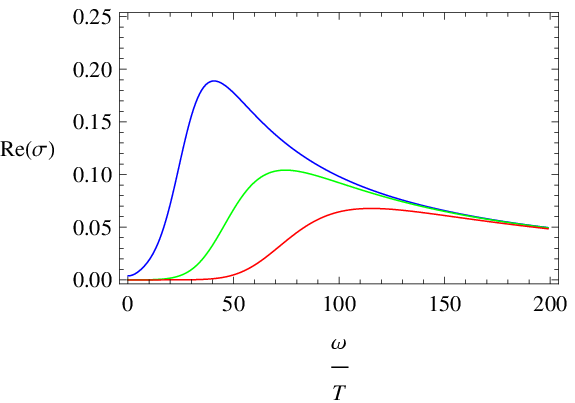}
\end{minipage}%
\begin{minipage}[c]{0.5\linewidth}
\includegraphics[width =2.8in,height=1.8in]{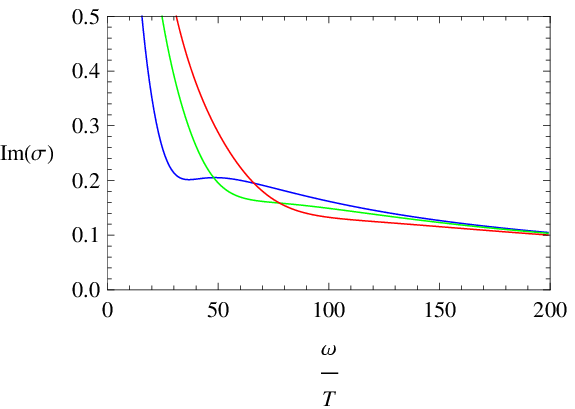}
\end{minipage}\quad
\begin{minipage}[c]{0.5\linewidth}
\includegraphics[width =2.8in,height=1.8in]{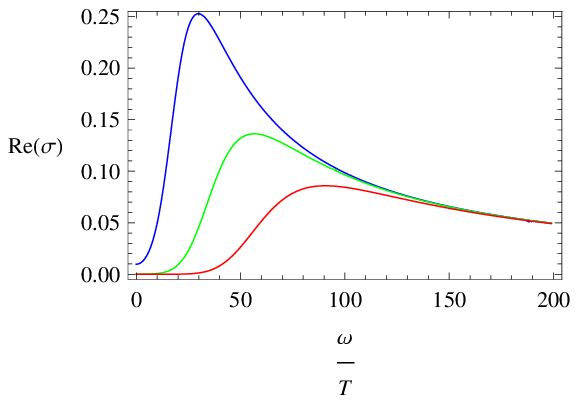}
\end{minipage}%
\begin{minipage}[c]{0.5\linewidth}
\includegraphics[width =2.8in,height=1.8in]{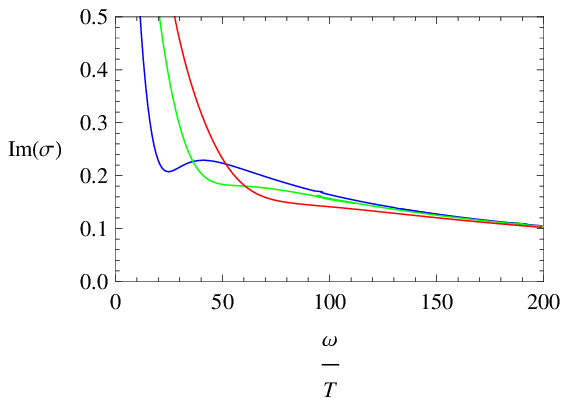}
\end{minipage}%
\caption{\label{fig:scalarcond2}Real and imaginary parts of the ac conductivity for the condensate ${\cal O}_{2}$ are plotted with respect to $\omega/T$ for different values of $J_r$ and $\alpha$. The top two graphs are  for ($J_r=2,~\alpha=0.001$) where the blue, green and the red curves correspond to different values of the temperature $T/T_c= 0.47, 0.27, 0.18$ respectively for both the real and the imaginary part of  the ac conductivity. The bottom two graphs are for ($J_r=0.02,~\alpha=0.001$) with the same values of the temperature.}
\end{figure}

Following a similar analysis as done in the case of the normal phase, the near horizon expansion of $A_{\varphi}$ with the incoming wave boundary condition may be expressed as,
\begin{equation}
A_{\varphi}(z)|_{z \rightarrow 1} = (f(z))^{-\frac{i \omega \sqrt{J_r + 2}}{2 \sqrt{2}}}(1+\cdots).
\end{equation}
Using the above equations we arrive at an expression for the ac conductivity in the superconducting phase in a similar fashion as that for the non rotating case (\ref{eq:AcCond}). In Fig.(\ref{fig:scalarcond1}) and Fig.(\ref{fig:scalarcond2}) we plot the real and the imaginary parts of the ac conductivity for various values of the parameters $J_r$ and $\alpha$.

The ac conductivity plots in Fig. (4) and Fig. (5) once again illustrates the dependence of both the real and imaginary parts of the conductivity on the angular momentum $J_r$ for the rotating charged BTZ black hole in our case. As emphasized earlier this is in sharp distinction to the non rotating case \cite {Ren} although the overall conductivity profiles are similar in our case. The origin for this dependence is expectedly the same as those described for the normal phase in Subsection 4.1. From both the figures (\ref{fig:scalarcond1}) and (\ref{fig:scalarcond2}) we observe that conductivity peaks are enhanced as we move from a higher value of $J_r$ to a lower value of $J_r$ i.e., from $J_r=2$ to $J_r=0.02$ for both the condensates ${\cal O}_{1}$ and ${\cal O}_{2}$.   The figures (\ref{fig:scalarcond1}) and (\ref{fig:scalarcond2}) clearly show that the peaks for the real and the imaginary parts of the ac conductivity for the condensate ${\cal O}_{2}$ are more pronounced than those for the condensate ${\cal O}_{
1}$. We also observe that the peaks are lowered for decreasing values of the temperature for both the condensates.

\section{Summary and Discussions}

In summary through the AdS$_3$/CFT$_2$ correspondence we have investigated both the normal and the superconducting phases of  a 1+1 dimensional boundary field theory on a circle $S^1$ which is holographically dual to a 2+1 dimensional bulk charged rotating BTZ black hole in the presence of a charged scalar field.  The ac conductivity of the boundary field theory has been computed numerically both for the normal and the superconducting phases of the boundary field theory in the probe limit and the low angular momentum approximation. For the normal phase we have established that the conductivity including the backreaction of the gauge fields, exhibits a dependence on the angular momentum $J$ of the rotating charged BTZ black hole in the bulk. The boundary field theory is expected to be described by some strongly coupled phase of a Fermi-Luttinger liquid on a circle as has been established in \cite {Bala,Nilanjan,Ren} and consequently the angular momentum $J$ is expected to be related to the interaction strength $g$  of the Fermi-Luttinger liquid.

In the superconducting phase of the boundary field theory we have shown that the bulk charged scalar field develops an instability below a certain critical temperature similar to the case of the non rotating charged BTZ black hole \cite {Ren}. This leads to the formation of  scalar hair for the rotating charged BTZ black hole in the bulk which corresponds to a charged condensate in the 1+1 dimensional boundary field theory on a circle. The formation of the charged condensate results in the spontaneous breaking of a global $U(1)$ symmetry in the boundary field theory and leads to a superconducting (superfluid)  phase transition. To this end we have implemented a careful recomputation of the analytic formulation described in \cite {Nurma} with attention to several missing terms, leading to a correct expression for the charged condensate in the superconducting phase of the boundary field theory.  The graphical description for the charged condensate in the boundary field theory following from the analytic formulation, have been augmented with those from a numerical computation for the same and compared. Both the analytic and the numerical results for the condensate compare favourably with the case of the non rotating charged BTZ black hole indicating a similar superconducting phase transition in the boundary field theory. We further determine a critical value of the angular momentum $J=J^{c}_{r}$ for fixed values of the chemical potential $\mu$, the temperature $T$ and the scalar field winding number $\alpha$ which serves as a lower bound for the superconducting phase transition in the boundary field theory. In the superconducting phase also the angular momentum $J$ and the scalar field winding $\alpha$ both relate to the coupling constant $g$ of the boundary field theory through the effective mass of the scalar field. The presence of a critical value of the angular momentum $J^c_r$  indicates a critical value $g_c$ for the interaction strength of the boundary Fermi-Luttinger liquid for the superconducting phase transition. 

Subsequently we have numerically computed the ac conductivity for the superconducting phase of the boundary field theory on the circle $S^1$ in the probe limit and the low angular momentum approximation. Both the real and imaginary parts of the ac conductivity exhibits a similar dependence on the angular momentum $J$ as in the normal phase which is in sharp distinction to those for the non rotating case in \cite {Ren}.  We mention in passing that the boundary field theory on a spatial circle described in our work is relevant for several interesting condensed matter physics applications such as mesoscopic rings and carbon nanotubes which are expected to be described by Luttinger liquids on a circle.

Our work leads to extremely interesting future directions for investigation. One of the possible avenues for this is to investigate the conductivity for 2+1 dimensional boundary theories on a sphere $S^2$ dual to a bulk charged rotating Kerr-Newman black hole in an AdS$_4$ space time along the lines of \cite {Sonner}. It would also be very interesting to relate our work to that in \cite {Bala} where a bulk charged rotating BTZ black hole in the presence of Wilson lines is related to helical Luttinger liquids at the boundary. From a condensed matter physics perspective it would be an interesting exercise to clearly understand the physics of the Fermi-Luttinger liquid in the context of our construction. We leave these interesting avenues for a future investigation.

\begin{center}
{\bf Acknowledgments}
\end{center}
We would like to thank the referee in Physical Review D for pointing out certain crucial issues leading to necessary modifications in our work. We thank S. Hartnoll and C. P. Hertzog for making their numerical codes publicly accessible. Useful discussions with Amit Agarwal, Sayantani Bhattacharya,  Amit Dutta, Tarun Kanti Ghosh, Tapobrata Sarkar and V. Subrahmanyam are gratefully acknowledged.   The work of Pankaj Chaturvedi is supported by Grant No. {\bf 09/092(0846)/2012-EMR-I} from the {\it Council of Scientific and Industrial Research  ( CSIR) }, India.

\end{document}